\documentclass[pra,onecolumn,nofootinbib,superscriptaddress]{revtex4}
\usepackage{calrsfs}
\usepackage{ulem}
\usepackage{graphics}
\usepackage{epstopdf}
\usepackage{bm}
\usepackage{bbm}
\usepackage{amssymb}
\usepackage{epstopdf}
\usepackage{graphicx}
\usepackage[caption=false]{subfig}
\usepackage{amsmath}
\usepackage{color}
\usepackage{array}

\begin{document}

\title{Quantum correlation light-field microscope with extreme depth of field}

\author{Yingwen \surname{Zhang}}
\email{yzhang6@uottawa.ca}
\affiliation{Nexus for Quantum Technologies, University of Ottawa, K1N 6N5, ON, Ottawa}
\affiliation{National Research Council of Canada, 100 Sussex Drive, Ottawa ON Canada, K1A0R6}

\author{Duncan \surname{England}}
\email{Duncan.England@nrc-cnrc.gc.ca}
\affiliation{National Research Council of Canada, 100 Sussex Drive, Ottawa ON Canada, K1A0R6}

\author{Antony \surname{Orth}}
\affiliation{National Research Council of Canada, 100 Sussex Drive, Ottawa ON Canada, K1A0R6}

\author{Ebrahim \surname{Karimi}}
\affiliation{Nexus for Quantum Technologies, University of Ottawa, K1N 6N5, ON, Ottawa}
\affiliation{National Research Council of Canada, 100 Sussex Drive, Ottawa ON Canada, K1A0R6}

\author{Benjamin \surname{Sussman}}
\affiliation{Nexus for Quantum Technologies, University of Ottawa, K1N 6N5, ON, Ottawa}
\affiliation{National Research Council of Canada, 100 Sussex Drive, Ottawa ON Canada, K1A0R6}

\begin{abstract}
Light-field microscopy (LFM) is a 3D microscopy technique whereby volumetric information of a sample is gained by simultaneously capturing both the position and momentum (angular) information of light illuminating a scene. Conventional LFM designs generally require a trade-off between position and momentum resolution, requiring one to sacrifice resolving power for increased depth of field (DOF) or vice versa. In this work, we demonstrate a LFM design that does not require this trade-off by utilizing the inherent correlations between spatial-temporal entangled photon pairs. Here, one photon from the pair is used to illuminate a sample from which the position information of the photon is captured directly by a camera. By virtue of the strong momentum anti-correlation between the two photons, the momentum information of the illumination photon can then be inferred by measuring the angle of its entangled partner on a different camera. By using a combination of ray-tracing and a Gerchberg-Saxton type algorithm for the light field reconstruction, we demonstrate that a resolving power of 5\,$\mu$m can be maintained with a DOF of $\sim500$\,$\mu$m, approximately 3 times of the latest LFM designs or $>100$ time that of a conventional microscope. In the extreme, at a resolving power of 100\,$\mu$m, it is possible to achieve near infinite DOF.
\end{abstract}	
\maketitle

\section{Introduction}

Utilizing the properties of quantum entangled photons to enhance the performance of sensing and imaging techniques has been an active area of research in recent decades. To date, most demonstrated techniques are focused on noise reduction in the low illumination regime by utilizing the strong correlations in one of the many degrees of freedom between entangled photon pairs, such as in time or position/momentum. Examples of this are sub-shot noise imaging~\cite{Brida2010,Samantaray2017} and quantum ghost imaging~\cite{Pittman1995,Shapiro2012,Padgett2016}. Dispersion immune optical coherence tomography has also been demonstrated using the two-photon Hong-Ou-Mandel interference effect~\cite{Nasr2003,Yepiz2022}. With advances in detector technology, especially with the maturity of single photon sensitive event cameras~\cite{Nomerotski2019,Defienne2021}, it is now possible to simultaneously utilize the correlations between multiple degrees of freedom. It has been demonstrated for further enhanced noise reduction in imaging and sensing by utilizing spectral-temporal correlations~\cite{Zhang2020} or spatial-temporal correlations~\cite{Defienne2021,Zhao2022} in entangled photons. New quantum imaging techniques utilizing the quantum correlations in multiple degrees of freedom have also been demonstrated, such as snapshot hyperspectral imaging~\cite{Zhang20222}, phase imaging through phase contrast microscopy~\cite{Hodgson2022} and Fourier Ptychography~\cite{Aidukas2019}, Hong-Ou-Mandel-microscopy~\cite{Ndagano2022}, as well as light-field/plenoptic imaging~\cite{Zhang2022}. In this work, we extend the quantum correlation light-field imaging design~\cite{Zhang2022} to microscopy and demonstrate volumetric reconstruction of a microscopic scene at the few micron resolution with extreme depth of field (DOF). 

Light-field or plenoptic imaging~\cite{Aldelson1992,Ng2005,raytrix} is a class of imaging techniques which allows for the reconstruction of the light field through capturing both the position and momentum (angular) information of light rays simultaneously. With the reconstructed light field, volumetric information of an illuminated scene can be obtained in a single measurement, with no scanning required. This technique has been since adapted to microscopy~\cite{Levoy2006,Bimber2019}, and has been demonstrated in volumetric non-scanning imaging of neural activities~\cite{Wang2022}, microendoscopy~\cite{Orth2019} and also shown to have potential applications in optogenetics~\cite{Schedl2016}. Conventional light-field microscope (LFM) designs typically make use of a microlens array (MLA). By placing the MLA one focal length away from an image sensor, each microlens illuminates a subset of the pixels in the CCD. By knowing which lens the light ray enters, and onto which pixel it subsequently focuses, one can obtain both position and momentum information of the light ray simultaneously, at the expense of sacrificing spatial resolution to gain momentum resolution. Typical LFM designs have each microlens covering around $10\times10$ pixels, thus resulting in a 10 times reduction to the position resolution and subsequently, resolving power. Increasing the size of each microlens to cover more pixels will increase the amount of momentum information captured and subsequently increase the DOF, however at the cost of further reduced position resolution. Techniques have been developed to improve either the DOF or resolving power such as through scanning the MLA~\cite{Lim2009} or the sample stage~\cite{Orth2012}. Others include applying 3D deconvolution~\cite{Broxton2013}, wavefront shaping~\cite{Cohen2014}, using a camera array~\cite{Lin2015}, processing the light-field information through the Fourier domain~\cite{Guo2019}, or using the aid of spherical aberrations in scanning LFM~\cite{Zhang20223}. 

In recent years, a new light-field imaging technique using spatially correlated thermal or quantum light has been proposed~\cite{DAngelo2016,Pepe2016,DiLena2018}. Here one beam/photon illuminates a scene to capture the position information, and the momentum information is inferred from the correlated partner beam/photon that is measured on a separate camera. Using this method, no trade-off between the position and momentum resolution is needed since each beam can be captured on separate cameras thus allowing for a much larger DOF while not sacrificing resolving power. This technique, has since been demonstrated using weakly correlated thermal light~\cite{Pepe2017,Massaro2022} and quantum-correlated light~\cite{Zhang2022}, with a DOF at around 10\,mm for a resolving power in the order of $100$\,$\mu$m. 

In this work, we present a proof-of-concept demonstration of light-field microscopy using quantum-correlated photons. We refer to this technique as quantum-correlation light-field microscopy (QCLFM) throughout the rest of the manuscript. Unlike the previous work~\cite{Zhang2022} that only employed ray optics, we utilize a wave optics approach in order to account for the diffraction effects that result from the much smaller target features. Our results demonstrate a resolving power of $5$\,$\mu$m (200\,lp/mm - line pairs per millimeter), with a depth of field (DOF) of $\sim500$\,$\mu$m. At a resolving power of $10$\,$\mu$m (100\,lp/mm), the DOF can be extended to over $1.5$\,mm. This DOF is $>100$ times that of a conventional microscope using the same objective lens. Compared to LFM designs, this is an order of magnitude larger compared to some of the earlier LFM demonstrations~\cite{Broxton2013,Cohen2014} and is $\sim3$ times larger than some of the most recent LFM techniques~\cite{Zhang20223}. In the extreme case, at a resolving power of $100$\,$\mu$m, we observed a nearly infinite DOF, a feat that is practically unfeasible with current LFM designs based on MLA. Limited by the single photon detection efficiency, the frame rate of QCLFM is currently in the order of 100\,s per frame, much slower compared to conventional LFM designs that operate at 10-100\,Hz. However, with expected advancements in single photon camera technology in the coming years, we are confident that QCLFM will become a potential alternative to conventional LFM designs.

\section{Results}
\subsection{Concept}
\begin{figure}
    \centering
    \includegraphics[width=1\linewidth]{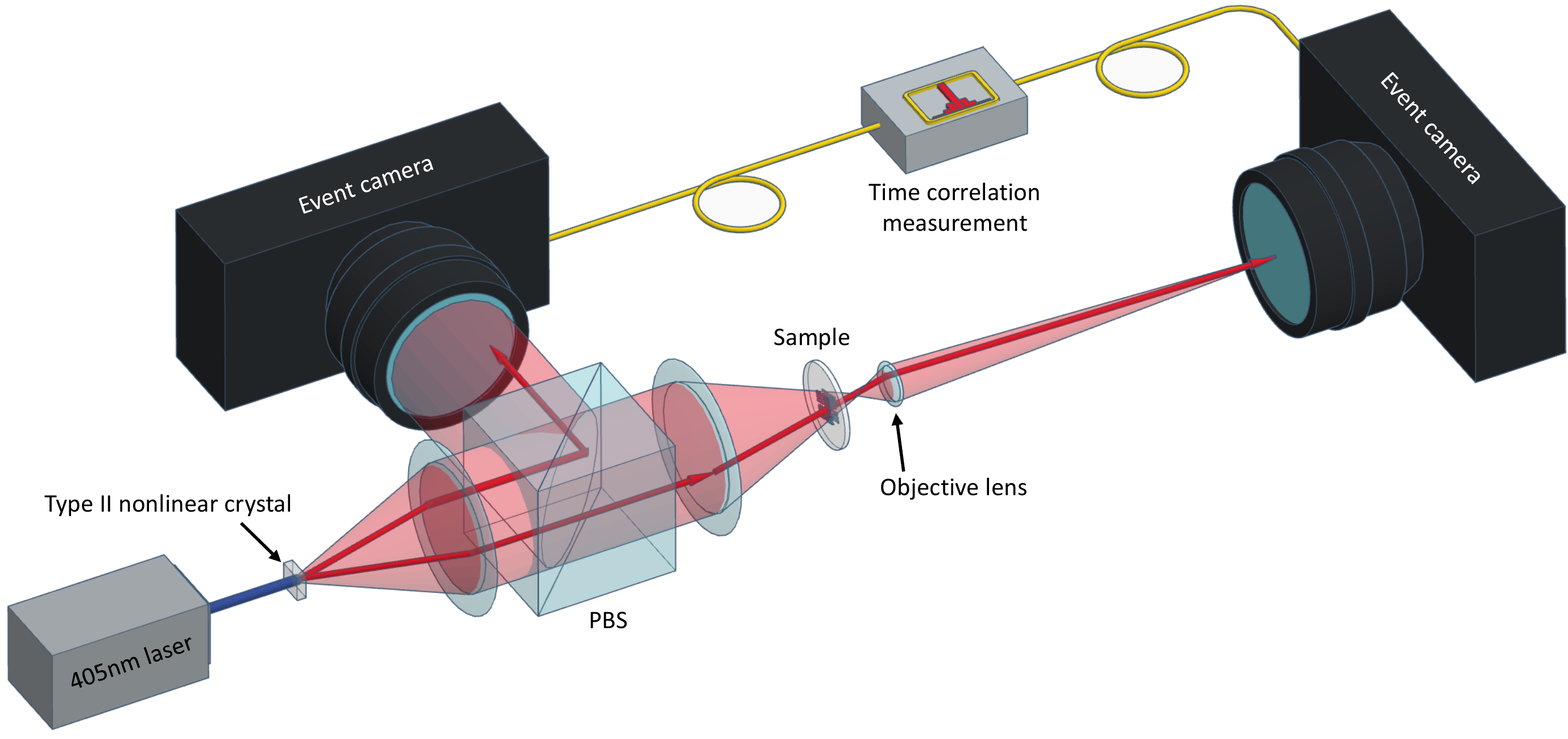}
    \caption{\textbf{Conceptual setup of the QCLFM system.} Pairs of photons with orthogonal polarization, correlated in time, position and momentum, are generated through the process of type II SPDC. The photon pairs, \textit{signal} and \textit{idler}, are separated into two separate paths via the use of a polarizing beamsplitter (PBS). In the signal path, the plane of the crystal is first imaged onto the sample placement region and then onto a time-tagging event camera where position information of the photons are captured. In the idler path, the Fourier plane of the nonlinear crystal is projected onto a different event camera to record the partner photons' momentum information. Time correlation measurements are then performed on all photons detected on the two cameras to identify the photon pairs. The momentum information of the signal photon can then be inferred from its time-correlated partner.}  
    \label{Fig1}
\end{figure}

The conceptual setup of the QCLFM system is shown in Fig.~\ref{Fig1}.  Through the process of type-II spontaneous parametric down-conversion (SPDC), a high energy photon from a UV pump laser is converted into a pair of lower energy photons in a nonlinear crystal. The photon pairs have orthogonal polarization and are entangled in time, position and momentum. It is this strong correlation property of entanglement that is used to realize QCLFM. The photon pairs, named \textit{signal} and \textit{idler} are first separated into two paths via the use of a polarizing beamsplitter (PBS). In the path of the signal photon, the plane of the crystal is first imaged onto the sample placement region and then onto a time-tagging event camera to record the signal photon's position information and arrival time. Through the path of the idler, the Fourier plane of the crystal is projected onto a different event camera to record the idler photons' momentum information and subsequent arrival time. The photon pairs are then identified through a time-correlation analysis based on their detection time and the signal photon's momentum information can then be inferred from its time-correlated partner.

\begin{figure}
    \centering
    \includegraphics[width=1\linewidth]{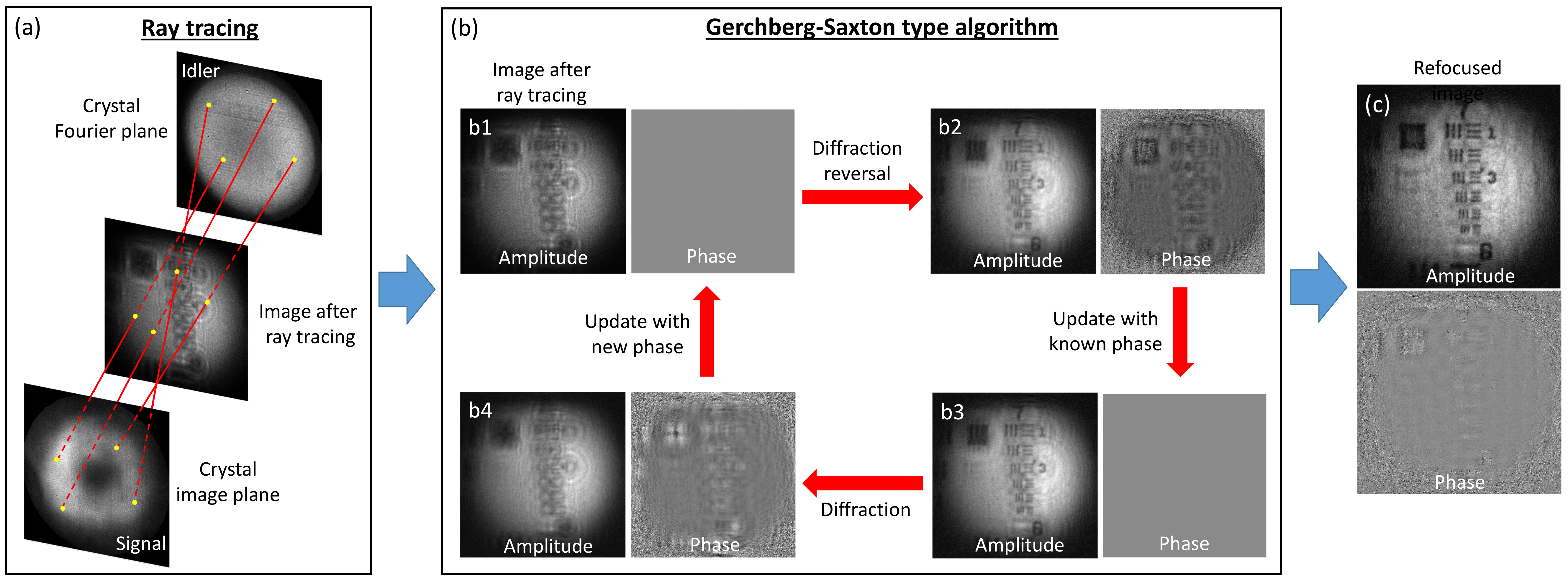}
    \caption{\textbf{Illustration of the digital refocusing process.} (a) Using the photon pairs locations detected on the nonlinear crystal's image plane and Fourier plane, a ray-tracing approach is first used to obtain the diffraction pattern amplitude of the sample as if illuminated by a plane wave propagating in the z-direction. (b) For a sample of uniform or known phase profile, a Gerchberg-Saxton type algorithm is used to recover the amplitude profile of the sample. (c) The recovered amplitude and phase of the sample after 10 loops in the algorithm. This illustration is that of a 1951 USAF resolution target (group 7) placed at $z=-300$\,$\mu$m away from the focus of the microscope objective.}  
    \label{Fig2}
\end{figure}

The operation for digital refocusing of a sample placed out of focus by a distance $z$ can be achieved using two steps. First, using the position and momentum information of each photon, and knowing the optical elements used between them, the trajectory of the photons can be reconstructed through a ray tracing operation. For large features, where diffraction effects are negligible, this first step, using ray optics, is enough to bring the sample back into focus~\cite{Zhang2022}, however, for smaller features, interference and diffraction effects from wave optics must also be taken into account. In this case, the image obtained after ray tracing is the diffraction pattern amplitude of the sample at a distance $z$ when illuminated by a plane wave propagating in the z-direction (see the Supplementary for details). The second step is to use a Gerchberg-Saxton type algorithm~\cite{Saxton1972,Fienup1982} to recover the amplitude of the sample. This refocusing process is illustrated in Fig.\ref{Fig2}. Details on the experimental setup and the refocusing procedure can be found in the Methods section.

\subsection{Experimental Results}
\begin{figure}
    \centering
    \includegraphics[width=1\linewidth]{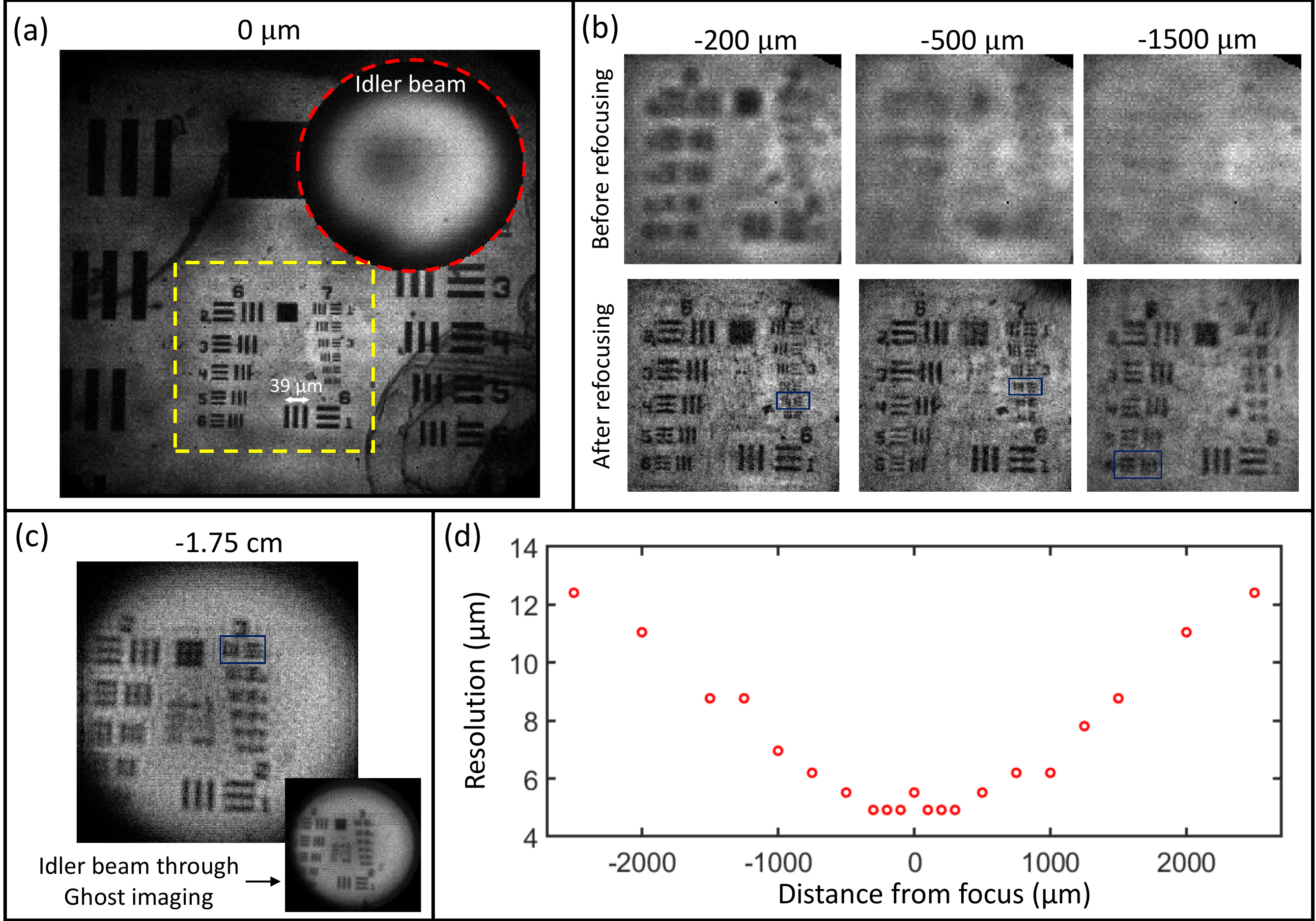}
    \caption{\textbf{Refocusing of 1951 USAF resolution target} (a) Image of a 1951 USAF resolution target, obtained after time correlation measurement, placed in the focus of a 20$\times$ 0.45\,NA objective lens. Here, for every photon detected in the idler beam, there is a signal photon detect in coincidence in the rest of the image. The region circled in red is the far field imaged by the idler photons. To not overshadow the rest of the image, the idler beam is only displayed at 30\% intensity.  (b) Before and after images of digitally refocusing groups 6 and 7 of a 1951 USAF resolution target (region in green square of (a)) placed at different distances from the objective focus with the smallest resolvable features after refocusing are boxed in blue. The data acquisition time for the images is 200\,s. (c) Digital refocusing of USAF target groups 2 and 3 placed 1.75\,cm away from the focus with a slightly different lens configuration. This is very close to the Fourier plane as the target can be clearly seen through ghost imaging in the idler beam (inset). See the main text for an explanation of ghost imaging. (d) Plot of the smallest resolution that can be refocused as a function of the distance a target is placed away from the focus with positive distance being towards the objective and negative away from the objective. The small peak at 0\,$\mu$m is a consequence of clustering in the raw camera data, see the Methods section for more details on this. For images of the full dataset of (b) and (c), see the Supplementary.} 
    \label{Fig3}
\end{figure}

For this experiment, only a single time-tagging event camera ($256\times256$ pixels with 55\,$\mu$m pixel pitch) is used. A corner of the camera of around $100\times100$ pixels was used for momentum measurement with the rest of the camera used for position measurement. With the momentum correlation of the photon pair measured to be approximately 2 pixels wide (13.6$\times10^{-3}$\,$\mu$m$^{-1}$/17.6$\times10^{-4}$\,rad), effectively around $50\times50$ pixels are used for momentum measurement. Though the idler beam is overlapped with part of the signal beam on the camera, since a coincidence measurement is performed, the noise in accidental coincidences from this overlap is negligible. This is based on the fact that the probability of having two signal photons generated within the coincidence gating time of 10\,ns is very low. See the Method section for more details.

In Fig.~\ref{Fig3} we show digitally refocused images of a 1951 USAF resolution target placed at different distances from the objective focus. We see that a line spacing of 5\,$\mu$m (group 7 element 5) can still be brought back into focus when the target is placed $\pm 500$\,$\mu$m away from the focus of the objective lens (positive being towards the objective lens and negative being away from) and for a line spacing of 10\,$\mu$m (group 6 element 5) this distance can be extended to more than 1500\,$\mu$m. In the extreme case, we are able to refocus a target placed at -1.75\,cm away from the focus with a line spacing of $\sim 100$\,$\mu$m, as seen in Fig.~\ref{Fig3}(c) which is very close to the Fourier plane, i.e. near infinity. This is verified through ghost imaging~\cite{Pittman1995,Shapiro2012,Padgett2016}, a correlation imaging technique using pairs of photons, in which the idler photon does not interact with the target being imaged, with only the signal photon interacting with the target. By time correlating the detection events of the two photons, an image of the object will appear through the idler photons even though they did not directly interact with the target. Knowing the idler beam is imaged in the crystal's Fourier plane, being able to see the target clearly after time correlation measurement is an indicator that the target is also placed at/near the Fourier plane.

Comparing to MLA based LFM demonstrations with a similar resolving power, we achieved around an order of magnitude improvement in DOF as compared to earlier LFM demonstration~\cite{Broxton2013,Cohen2014} (group 7 element 3 with 1000\,$\mu$m DOF compared 100\,$\mu$m DOF), and $\sim$3 times improvement in DOF compared to one of the latest LFM demonstrations~\cite{Zhang20223} (group 7 element 3 with 1000\,$\mu$m DOF compared to 300\,$\mu$m DOF and group 6 element 3 with 2000\,$\mu$m DOF compared to 800\,$\mu$m DOF), all of which uses more sophisticated techniques such as deconvolution, wavefront shaping and spherical aberration assistance for improving position resolution and DOF. In addition, we see that QCLFM is also insensitive to the direction at which the target is placed from the objective focus as compared to some conventional LFM techniques~\cite{Broxton2013,Zhang20223}. On the other hand, a conventional microscope with an objective numerical aperture of 0.45, as used in this experiment, would have a depth of field of approximately $5$\,$\mu$m (see Supplementary materials for details). 

\begin{figure}
    \centering
    \includegraphics[width=1\linewidth]{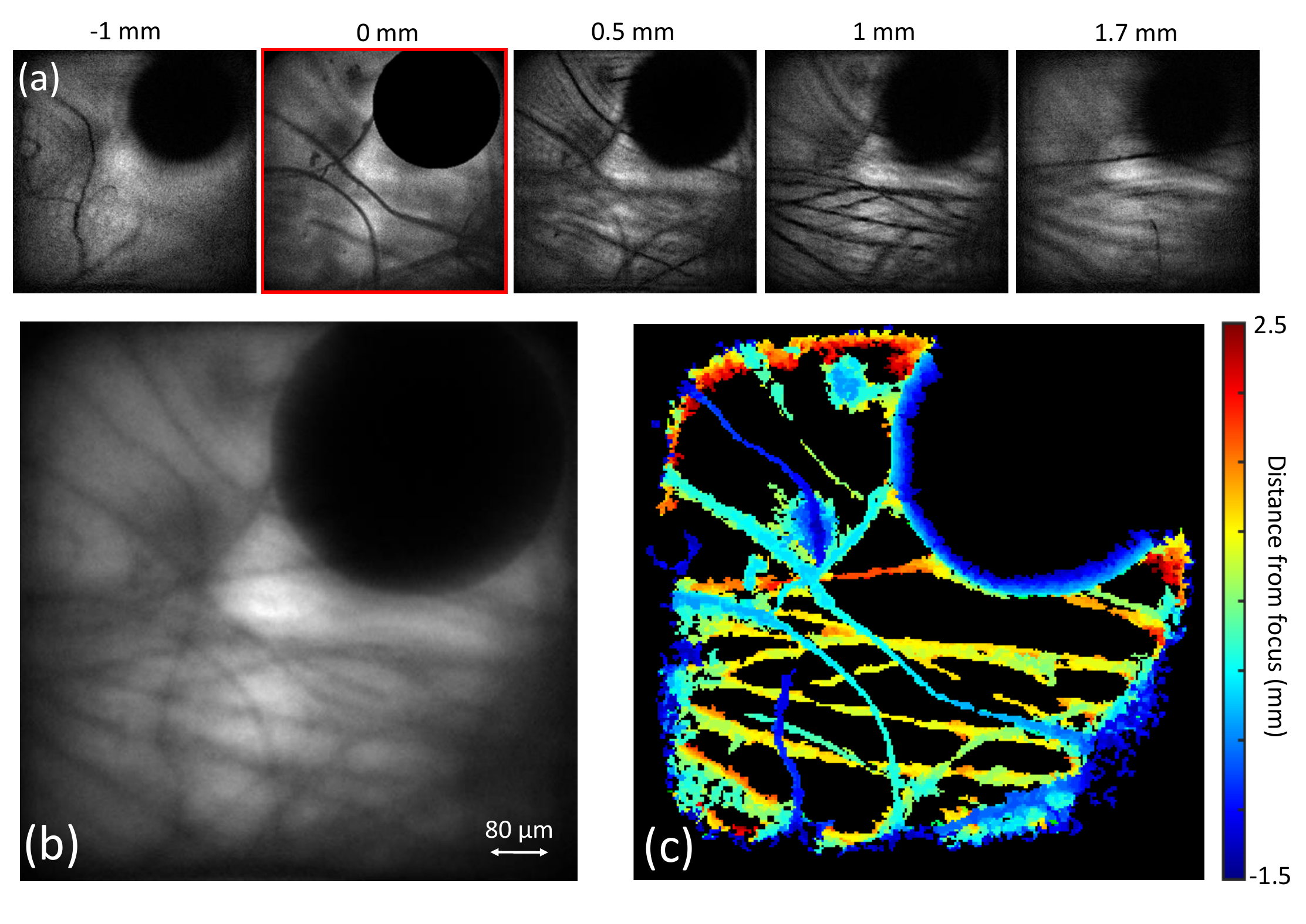}
    \caption{\textbf{Refocusing of lens tissue fiber} (a) Digitally refocusing onto fiber strands of lens-cleaning tissue at various distances from the focus of the objective lens. The image at the lens focus is highlighted in red. The fiber strands are from the edges of a $\sim 3$\,mm thick stack of lens cleaning tissue. The measurement is taken in with a data acquisition time of 200\,s. The empty circular region on the top right is the part of the camera used for capturing the idler photons. (b) Image showing all fiber strands brought into focus through post-processing. (c) Depth map of the fiber strands. A video showing the refocused images at different depths can be found in the Supplementary.} 
    \label{Fig4}
\end{figure}

Next, we test our technique on a more complex 3D scene. Imaging of multiple overlapping fiber strands, between $5-10$\,$\mu$m in diameter, on the edges of a $\sim 3$\,mm thick stack of lens-cleaning tissue can be seen in Fig.~\ref{Fig4}. A single measurement was taken on the stack of lens tissue with digital refocusing performed in post-processing. We can see fiber strands more than 1\,mm away from the objective lens focus, and completely not visible at the objective focus, being brought back into focus. The post-processing results in a stack of in-focus images obtained at various depths in the sample, some examples of which are shown in Fig.~\ref{Fig4}(a). The sum of all of these depth images results in an ``all in focus" image (Fig.~\ref{Fig4}(b)) in which every strand, regardless of its position, is bought into focus. Finally, by colour-coding each image based on its depth, we generate a depth map of the 3D scene shown in Fig.~\ref{Fig4}(c).

Some artifacts/halos can seen in the refocused images, this is likely a result of wavefront distortions from the multiple lenses used in the setup which were not taken into account during the refocusing procedure. Another likely source is in the approximations made in the refocusing procedure, such as small angle approximations and the uniform illumination beam intensity (see the Supplementary for details). 

\section{Discussion}
We have made a proof-of-concept demonstration of a LFM design based on utilizing the inherent position and momentum correlation of entangled photon pairs generated through the process of SPDC. Since each degree of freedom can potentially be measured on separate cameras, one does not need to sacrifice position resolution for momentum resolution or vice versa as in conventional LFM designs. This has allowed us to achieve a DOF that is approximately 3 times larger than that of the latest LFM designs, all at an illumination power of about 3\,pW ($15\times10^6$ photons per second at 810\,nm). With the high momentum resolution (effectively $\sim 50\times50$ pixels at $13.8$\,$\mu$m$^{-1}$ per pixel), it is even possible to digitally refocus objects placed at near infinity, though with a moderate resolving power. In addition, as compared to some conventional LFM techniques, QCLFM is also insensitive to whether the target is placed before or after the objective focus.

The DOF can be further increased if a higher degree of momentum correlation is achieved with a separate camera used for the momentum measurement on the idler photon. Momentum correlation can be improved by increasing the pump beam waist or using a pump with a longer coherence length. Details on this can be found in the Methods section. Position resolution can be further improved by a factor of 3 on the event camera (TPX3CAM) using centroiding techniques~\cite{Kim2020}. QCLFM only utilizes the strong correlations inherent in spontaneous parametric downconversion, and does not leverage photon entanglement. Thus it is possible to imagine combining QCLFM with other quantum sensing techniques which directly utilize entanglement, such as quantum lithography~\cite{Boto2000,DAngelo2001}, to achieve resolution beyond the diffraction limit. Additionally, the current QCLFM design is still in its most basic form, we believe significant performance gains can be made by incorporating more sophisticated techniques already developed for the conventional LFM design such as 3D deconvolution and wavefront shaping. 

One limitation of the current QCLFM design and refocusing algorithm is that it can only be used for targets with a known or constant phase profiles. A potential method to solve this is to also capture the beam in an intermediate plane in addition to just the near and far field. With this extra information one may be able to determine both the amplitude and phase of the diffraction pattern of the sample at two different planes from which the phase and amplitude of the sample can be retrieved. The feasibility of this approach will require further investigation. Another limitation to the technique's performance (and all single photon quantum imaging techniques in general) is in its slow data acquisition speed, which is in the order of a few hundred seconds per frame as compared to the 10-100\,Hz frame rate of classical LFM designs. This is mainly limited by the detection efficiency and timing resolution of the detection camera. The single photon detection efficiency of the camera used here is $\sim7\%$ with an effective timing resolution of $\sim8$\,ns~\cite{Vidyapin2022}. Since the pair detection efficiency is quadratic with respect to the single photon detection efficiency, with a camera efficiency of 50$\%$, we could expect a $(50/7)^2\approx50$ times improvement in the data acquisition speed. Moreover, the signal-to-noise ratio of temporal correlation measurement scales as the square-root of the timing resolution, so if a timing resolution of 0.1\,ns can be achieved, a further 10 times reduction to the data acquisition time is possible. Single-photon avalanche photodiode array cameras meeting these specifications are currently in development and may become commercially available soon~\cite{Canon}. We expect this makes real time imaging and microscopy with QCLFM (and many other quantum imaging techniques) a possibility in the near future.

\section{Methods}
\subsection*{Experimental Setup}
\begin{figure}
    \centering
    \includegraphics[width=1\linewidth]{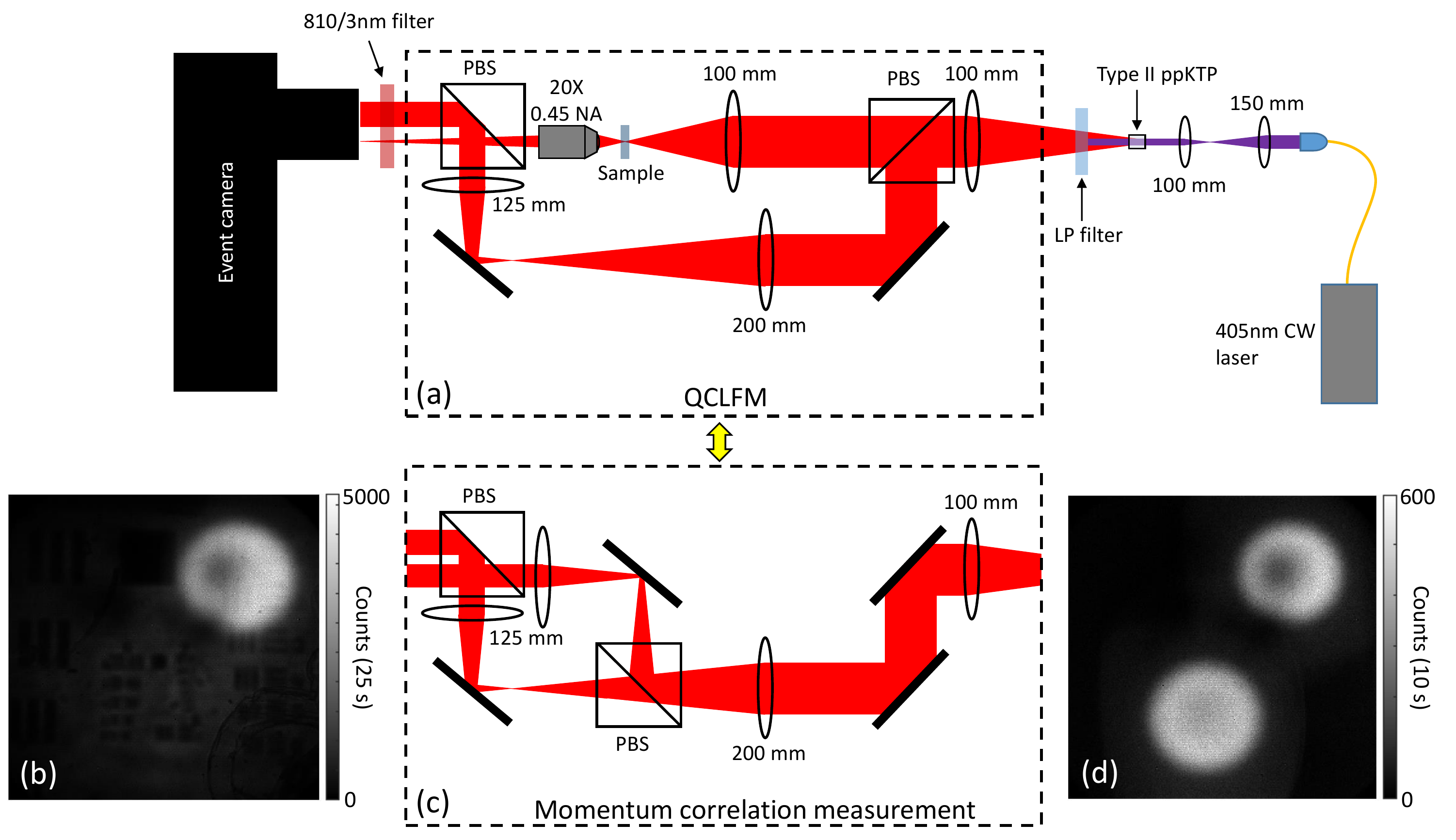}
    \caption{\textbf{Experimental setup of QCLFM} (a) Experimental setup of the QCLFM. PBS: polarizing beamsplitter, LP filter: longpass filter. (b) A typical image captured on the camera, before time correlation analysis, accumulated over 25\,s, of a 1951 USAF resolution target placed 200\,um away from the objective focus. The signal beam covers the whole sensor area, while the idler is the bright region on the top-right corner. (c) The modified experimental setup for performing momentum correlation measurement. (d) Image captured on the camera, before time correlation analysis, accumulated over 10\,s, for performing momentum correlation measurement. The darker patch seen in the idler beam in both (b) and (d) is due to a damaged region on the camera system resulting in lower detection efficiency.}
    \label{Methods1}
\end{figure}

The schematic of our QCLFM setup is shown in Fig.~\ref{Methods1}(a). A fiber coupled 405\,nm continuous wave (CW) laser, at $\sim30$\,mW, is used to pump a $1\times2\times1$\,mm (H$\times$W$\times$L) Type~II periodically-poled potassium titanyl phosphate (ppKTP) crystal to produce entangled photon pairs through SPDC, at a rate of approximately $15\times10^6$ photon pairs per second or, after accounting for the $7\%$ camera detection efficiency, about 15 photons per second per pixel detected on the camera. The photon pairs are orthogonal in polarization, correlated in position, momentum and time. The SPDC photons are then separated by a polarizing beam splitter (PBS) into two paths. In the path of the signal photon, the plane of the crystal is first imaged onto the sample placement region and then onto a time-tagging event camera (TPX3CAM~\cite{Nomerotski2019,ASI}) to record the signal photon's position information and arrival time. In the path of the idler photon, the Fourier plane of the crystal is projected onto a corner of the camera to record the idler photons' momentum information and the subsequent arrival time. Thus, the position information of the signal photon is captured directly and its momentum information can be inferred from the idler partner. The photon pairs are identified through a time correlation measurement, with a coincidence gating time of 10\,ns, on all photons detected between the idler beam region and the rest of the camera. A typical image captured, before time correlation analysis, can be seen in Fig.~\ref{Methods1}(b). The idler beam on the corner is seen much brighter compared to the signal beam covering the whole sensor, this is due to a similar amount of photons being concentrated onto a smaller area. 

The TPX3CAM has $256\times256$ pixels with a pixel pitch of 55\,$\mu$m. The pixels can individually time the arrival time of a light pulse with 1.6\,ns accuracy. The camera is made single photon sensitive with an attached image intensifier in which a single photon is converted into a flash of light bright enough to be registered by the camera. The flash of light will illuminate a small cluster of pixels on the camera, on which a clustering identification and centroiding algorithm must be implemented to regroup each cluster into a single event. A consequence of the clustering is a slight blurring of the raw image data, resulting in it being difficult to find the objective focus precisely. This is what resulted in the slight decrease in resolution at 0\,$\mu$m seen in Fig.~\ref{Fig3}(d). The timing of the camera pixels is intensity dependent, in which brighter pixels of a cluster will be registered as an earlier event than dimmer pixels of the same cluster, even though they are created from the same single photon event. This intensity-dependent timing must also be corrected and will result in losing some timing accuracy giving an effective timing accuracy of $\sim 8$\,ns. The quantum efficiency of the camera system (camera + intensifier) was measured to be $\sim 7\%$. More details on the camera and the post-processing required can be found in~\cite{Zhao2017,Vidyapin2022}.  

The lenses in the signal beam were chosen such that the signal beam is not too tightly focused onto the sample region, allowing the field of view to stay roughly constant in the region of $\sim\pm 2$\,mm from the objective focus. When magnified onto the camera, the beam size should also roughly match the sensor area of the camera to ensure most photons will be captured. Likewise, in the idler beam, the lenses were chosen such that the idler beam will cover roughly a region of $100\times100$ pixels of the camera. 

In order to achieve the highest degree of momentum correlation possible, a pair of lenses with focal length 100\,mm and 150\,mm were used to collimate and demagnify the laser beam to $\sim 1$\,mm in diameter when hitting the ppKTP crystal. The expected degree of momentum correlation can be calculated as~\cite{Defienne2019}
\begin{equation}
    \sigma_{k_1+k_2} = \sqrt{1/l_c^2+1/(4\omega_p^2)},
    \label{kcorr}
\end{equation}
where $\sigma_{k_1+k_2}$ is the standard-deviation in the momentum correlation between two photons with momentum $k_1$ and $k_2$, $l_c$ is the coherence length of the pump laser and $\omega_p$ is the pump beam waist radius at the crystal. For our pump laser, $l_c\approx 200$\,$\mu$m and $\omega_p\approx 500$\,$\mu$m, thus giving $\sigma_{k_1+k_2}\approx 5\times10^{-3}$\,$\mu$m$^{-1}$. 

The experimental setup to perform momentum correlation measurement is shown in Fig.~\ref{Methods1}(b). Here, the Fourier plane of the crystal is imaged by both the signal and idler photons, each onto a separate region of the camera. The common 200\,mm lens is moved along the beam path until the highest degree of momentum correlation is observed. This setup is designed such that alteration of the idler path is kept to a minimum when converting to the QCLFM setup of Fig.~\ref{Methods1}(a). Momentum correlation measurement should be performed first in order to ensure that when converted to QCLFM, the camera is imaging the exact Fourier plane of the crystal which would contain the highest degree of momentum correlation.

\begin{figure}
    \centering
    \includegraphics[width=1\linewidth]{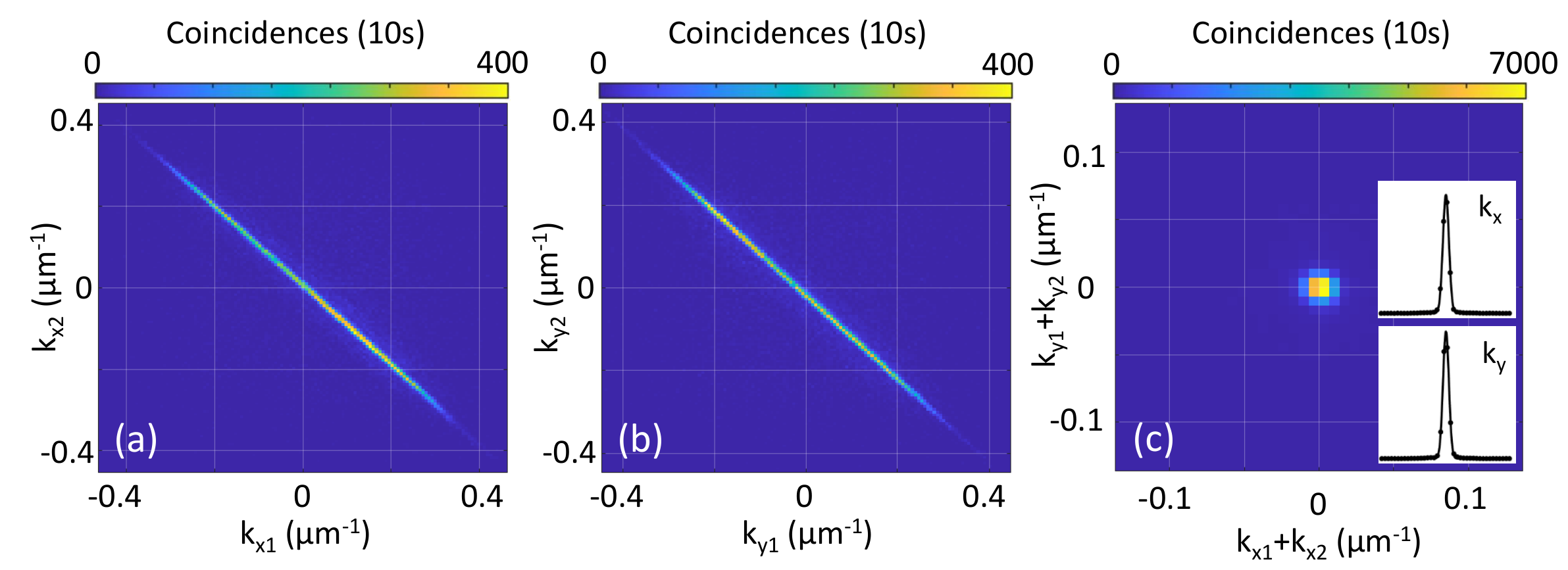}
    \caption{\textbf{Momentum correlation measurement} (a) Joint probability distribution of photon momentum in the x-direction $k_x$. (b) Joint probability distribution of photon momentum in the y-direction $k_y$.  (c) Sum coordinate projection of the joint probability distribution with the fitted Gaussian to the cross-section of the $k_x$ and $k_y$ direction shown in the inset. Fitted Gaussian gives $\sigma_{k_{x1}+k_{x2}} = 6.1\times10^{-3}$\,$\mu$m$^{-1}$ (0.90\,pixels) and $\sigma_{k_{y1}+k_{y2}} = 6.4\times10^{-3}$\,$\mu$m$^{-1}$ (0.93\,pixels). }
    \label{Methods2}
\end{figure}

The measured momentum correlation, determined by fitting a Gaussian of the form $f(k) = a \exp\left(-\frac{(k-b)^2}{2\sigma_{k_1+k_2}^2}\right)$ to the central peak of the sum coordinate projection of the momentum joint probability distribution seen in Fig.~\ref{Methods2}(c), is $\sigma_{k_{x1}+k_{x2}} = 6.1\times10^{-3}$\,$\mu$m$^{-1}$ (0.90\,pixels) and $\sigma_{k_{y1}+k_{y2}} = 6.4\times10^{-3}$\,$\mu$m$^{-1}$ (0.93\,pixels) in the x and y direction respectively. This is in agreement with the expected $5\times10^{-3}$\,$\mu$m$^{-1}$ as calculated using Eq.\,(\ref{kcorr}). 

The achievable DOF of the current experimental design is mostly limited by the degree of momentum correlation. Although approximately $100\times100$ pixels have been allocated for momentum measurement, effectively, the momentum resolution is only $50\times50$ as the FWHM of the momentum correlations is 2.1 pixels wide. Using a laser with a longer coherence length of $l_c = 2$\,mm will be able to improve $\sigma_{k_1+k_2}$ by a factor of 5 which would allow the full sensor area of the camera ($256\times256$ pixels) to be effectively used for momentum measurement when two cameras are used as the experiment was initially designed. 

\subsection*{Digital Refocusing Procedure}

Using the position and momentum information of each photon, the operation for digital refocusing of a sample placed out of focus by a distance $z$ can be achieved using two steps. First, the trajectory of the photons are determined through ray tracing using a ray transfer matrix
\begin{equation}
    \begin{bmatrix}
    \overrightarrow{r_2} \\
    \overrightarrow{\theta_2} 
    \end{bmatrix}
    =
    \begin{bmatrix}
    A & B \\
    C & D
    \end{bmatrix}    
    \begin{bmatrix}
    \overrightarrow{r_1} \\
    \overrightarrow{\theta_1} 
    \end{bmatrix},
\label{ABCDmatrix}
\end{equation}
where $\overrightarrow{r_1}$ and $\overrightarrow{r_2}$ are the positions of each photon pair detected at the two camera image planes, $\overrightarrow{\theta_1}$ and $\overrightarrow{\theta_2}$ are the angles at which the photons hit each plane and the ray transfer (ABCD) matrix is determined by the optical components placed between the two image planes. Since $\overrightarrow{r_1}$ and $\overrightarrow{r_2}$ are measured directly by the two cameras and the ABCD matrix is also known, $\overrightarrow{\theta_1}$ and $\overrightarrow{\theta_2}$ can be determined from Eq.~(\ref{ABCDmatrix}). Knowing $\overrightarrow{r}$ and $\overrightarrow{\theta}$, an appropriate ABCD matrix is then used to digitally propagate each photon to the location of the sample. The image obtained after this first step is the sample's diffraction pattern amplitude at a distance $z$ when illuminated by a plane wave propagating in the z-direction (this is theoretically shown in the Supplementary Materials). 

Next, a Gerchberg-Saxton type algorithm can be used to recover the amplitude of the sample if the sample has a uniform or known phase profile. The traditional Gerchberg-Saxton algorithm~\cite{Saxton1972,Fienup1982} uses the amplitude information obtained at the image plane and Fourier plane to retrieve the phase information of the two through the Fourier transform relation between the two planes. Here, we use the amplitude of an arbitrary plane and the phase of another to retrieve the corresponding missing phase and amplitude of the two planes. The relationship between the two planes is no longer a direct Fourier transform but that of wave diffraction where techniques for approximating wave diffraction, such as the angular spectrum method or Fresnel diffraction method, must be used. The algorithm works as follows:
\begin{enumerate}
    \item Guess an initial phase profile for the diffraction pattern amplitude obtained after ray tracing (a flat initial phase was used here).
    \item Apply the angular spectrum method to the diffraction pattern to reverse diffraction and obtain an amplitude and phase profile of the sample (Fig.\,\ref{Fig2}~b1 to b2).
    \item Replace the phase profile of the sample obtained from 1. with the known phase profile (which is zero for the targets used in this experiment) and then apply the angular spectrum method to obtain the amplitude and phase of its diffraction pattern (Fig.\,\ref{Fig2}~b2 to b3).
    \item Update the diffraction pattern obtained from ray tracing with the new phase profile (Fig.\,\ref{Fig2}~b3 to b4).
    \item Repeat the process from step 2.
\end{enumerate}

The angular spectrum method is applied as follows, for a beam with an initial amplitude and phase profile of $U_i(x,y)e^{i\phi_i(x,y)}$, the profile after propagating a distance $z$ is
\begin{equation}
    U_f(x,y)e^{i\phi_f(x,y)} =  \mathcal{F}^{-1}\left[\mathcal{F}\left(U_i(x,y)e^{i\phi_i(x,y)}\right)e^{ik_z z}\right],
\end{equation}
where $\mathcal{F}$ ($\mathcal{F}^{-1}$) denotes the (inverse) Fourier transform operation and $k_z = \sqrt{k^2-k_x^2-k_y^2}$ with $k = 2\pi/\lambda$ being the photon wavenumber.

\section*{Acknowledgements}
The authors are grateful to Denis Guay, and Doug Moffatt for technical support. The authors acknowledge the support of Canada Research Chair, NRC-uOttawa Joint Centre for Extreme Quantum Photonics (JCEP), Quantum Sensors Challenge Program at the National Research Council of Canada and Defence Research and Development Canada. 

\bibliographystyle{unsrt}
\bibliography{LFMref}

\section*{Supplementary Materials} 

\subsection*{Refocusing Operation}

For a plane wave traveling in an arbitrary direction $U(x,y,z)=Ae^{i(ux+vy+wz)}$ (with $2\pi/\lambda = k = \sqrt{u^2+v^2+w^2}$) illuminating a target with complex spatial profile $f(x,y) = \mathcal{F}^{-1}\left[F(k_x/2\pi,k_y/2\pi)\right]$ ($\mathcal{F}$ denoting the Fourier transform operation) placed at $z=0$, the transmitted wave can then be written as a superposition of plane waves (also known as the angular spectrum method) given by
\begin{align}
    U_T(x,y,z) &= \frac{1}{4\pi^2} \int\int^{\infty}_{-\infty} F(k_x/2\pi,k_y/2\pi) e^{i(k_xx+k_yy)}U(x,y,0)e^{i\tilde{k}_zz}dk_xdk_y \nonumber\\
    &= \mathcal{F}^{-1}\left[F(k_x/2\pi,k_y/2\pi) e^{i(\tilde{k}_zz)}\right]U(x,y,0),
    \label{eq1}
\end{align}
with $\tilde{k}_z = \sqrt{k^2 -(k_x+u)^2 - (k_y+v)^2}$.

In the case of small angles approximation where the x and y components of the k-vector is much smaller than the z-component, we can approximate $\tilde{k}_z$ using a Taylor series expansion
\begin{align}
    \tilde{k}_z &= \sqrt{k^2 - k_x^2 - k_y^2 - 2k_xu - 2k_yv - u^2 - v^2}\nonumber\\
                &\approx k_z -\frac{k_xu+k_yv}{k_z} + O(u^2,v^2),
\end{align}
where $k_z = \sqrt{k^2 -k_x^2 - k_y^2}$.

Now substituting back into eq.\ref{eq1} we have
\begin{align}
    U_T(x,y,z) &\approx \mathcal{F}^{-1}\left[F(k_x/2\pi,k_y/2\pi) e^{i\left(k_z -\frac{k_xu+k_yv}{k_z}\right)z}\right] U(x,y,0)\nonumber\\
    &= \mathcal{F}^{-1}\left\{\mathcal{F}\left[f(x-\Delta x,y-\Delta y)\right] e^{ik_zz}\right\}U(x,y,0)\nonumber\\
    &= \tilde{f}(x-\Delta x,y-\Delta y,z)U(x,y,0),
   \label{eq3}
\end{align}
where we have used the Fourier transform property $\mathcal{F}\left[f(x-\Delta x\right] = F(k_x/2\pi)e^{-ik_x\Delta x}$ with $\Delta x = \frac{uz}{k_z}$ and $\Delta y = \frac{vz}{k_z}$ and $\tilde{f}(x,y,z) \equiv \mathcal{F}^{-1}\left\{\mathcal{F}\left[f(x,y)\right] e^{ik_zz}\right\}$ is the diffracted field of a target $f(x,y)$ illuminated by a plane wave traveling in the z-direction. 

For this experiment, there are an infinite number of plane waves illuminating the target of which we are able to measure a total of $N \times M$, the total number of pixels used to capture the idler photons in the far-field plane, Thus we will denote the transmitted wave associated with a particular far-field pixel, with pixel index $n$ and $m$ in the x and y direction respectively, as $U_T{_{nm}}(x,y,z) = \tilde{f}(x-\Delta x_n,y-\Delta y_m,z)U_{nm}(x,y,0)$ with $U_{nm}(x,y,0) = A_{nm}e^{i(u_nx+v_my)}$. The captured image in the near-field plane $I_0(x,y,z_0)$ is thus given by the sum of all the transmitted waves squared, i.e.
\begin{equation} 
    I_0(x,y,z_0) = \left|\sum_{n,m} \tilde{f}(x-\Delta x_n,y-\Delta y_m,z_0)U_{nm}(x,y,0)\right|^2.
    \label{eq4}
\end{equation}

Since we can track each photon in forming $I_0(x,y,z_0)$ from the far-field to the near-field plane, we can break up $I_0(x,y,z_0)$ into a total of $N\times M$ individual images $\tilde{I}_{nm}(x,y,z_0)$ for each $(u_n, v_m)$ pixel in the far-field thus
\begin{equation} 
    I_0(x,y,z_0) = \sum_{n,m} \tilde{I}_{nm}(x,y,z_0) = \left|\sum_{n,m} \tilde{f}(x-\Delta x_n,y-\Delta y_m,z_0)U_{nm}(x,y,0)\right|^2.
\end{equation}
Making a coordinate shift gives
\begin{align}
    \sum_{n,m} \tilde{I}_{nm}(x+\Delta x_n,y+\Delta y_m,z_0) &= \left|\sum_{n,m} \tilde{f}(x,y,z_0)U_{nm}(x+\Delta x_n,y+\Delta y_m,0)\right|^2\nonumber\\
    &= \left|\tilde{f}(x,y,z_0)\right|^2 \left|\sum_{n,m}U_{nm}(x+\Delta x_n,y+\Delta y_m,0)\right|^2.
    \label{eq6}
\end{align}

On the left side of eq.\,\ref{eq6}, the operation of shifting each image $\tilde{I}_{nm}$ by a distance of $\Delta x_n = \frac{u_nz_0}{k_z}$ and $\Delta y_m = \frac{v_mz_0}{k_z}$ in the x-y plane is equivalent to the ray-tracing operation of shifting the position of each photon by the same factor based on the direction the photon is traveling given by $\theta_x \approx  \frac{u_n}{k_z}$  and $\theta_y \approx  \frac{v_m}{k_z}$ after propagating a distance $z_0$. On the right hand side, the first term is the intensity of the diffraction pattern of the target $\left|\tilde{f}(x,y,z_0)\right|^2$ as if obtained when illuminated by a plane wave traveling in the z-direction. The second term is the intensity of the illuminating beam at $z=0$, $\left|\sum_{n,m}U_{nm}(x+\Delta x_n,y+\Delta y_m,0)\right|^2$, with each plane wave also shifted by $\Delta x_n$ and $\Delta y_m$ in the x-y plane. If the intensity of the illuminating beam is uniform across the target then, 
\begin{equation} 
    \sum_{n,m} \tilde{I}_{nm}(x+\Delta x_n,y+\Delta y_m,z_0) \propto \left|\tilde{f}(x,y,z_0)\right|^2.
\end{equation}

Thus, with the amplitude of the diffraction pattern known and if we also know the phase profile of the target, the amplitude of the target can be retrieved using a Gerchberg-Saxton type algorithm whose implementation is detailed in the main text.

\section{Depth of field of a conventional microscope}

The depth of field for a conventional optical microscope is given by
\begin{equation}
    \text{DOF} = \frac{n \cdot \lambda}{\text{NA}^2} + \frac{n\cdot e}{\text{M}\cdot\text{NA}},
\end{equation}
where, for this experiment, $n=1$ is the refractive index in air, $\lambda=810$\,nm is the photon wavelength, $\text{NA}=0.45$ is the objective numerical aperture, $\text{M}=20$ is the magnification and $e$ the distance resolved. So for $e=5$\,$\mu$m the DOF is 4.6\,$\mu$m and for $e=10$\,$\mu$m the DOF is 5.1\,$\mu$m, thus QCLFM has a DOF that is at least 2 orders of magnitude larger compared to a conventional microscope.

\subsection*{Images of full dataset }
All images for the dataset shown in Fig.~3(d) of the main manuscript is shown below in Fig.~\ref{Supp1}.

Fig.~\ref{Supp2} shows the dataset when using a 35\,mm focal length condenser lens instead of the 100\,mm condenser lens as used for Fig.~\ref{Supp1}. This configuration was chosen such that the smallest resolvable feature at each distance is close to the resolution limit of the camera. With this configuration, at the objective focus, the field of view (FOV) is approximately 1/3 of that when using a 100\,mm condenser lens. However, the FOV expands much more rapidly with the 35\,mm lens allowing larger target feature to better fit into the FOV when far away from the objective focus.

In Fig.~\ref{Supp2}, we can see that beyond -200\,$\mu$m away from the objective focus, the presence of the target can no longer be observed in the near field of the crystal plane imaged by the signal beam. From -2.8\,mm away from the objective focus, the presence of the target can once again be observed through ghost imaging in the far field of the crystal plane imaged by the idler beam. A plot showing the smallest resolvable features vs. the distance from focus is shown in Fig.~\ref{Supp3}. 

\begin{figure}
    \centering
    \includegraphics[width=1\linewidth]{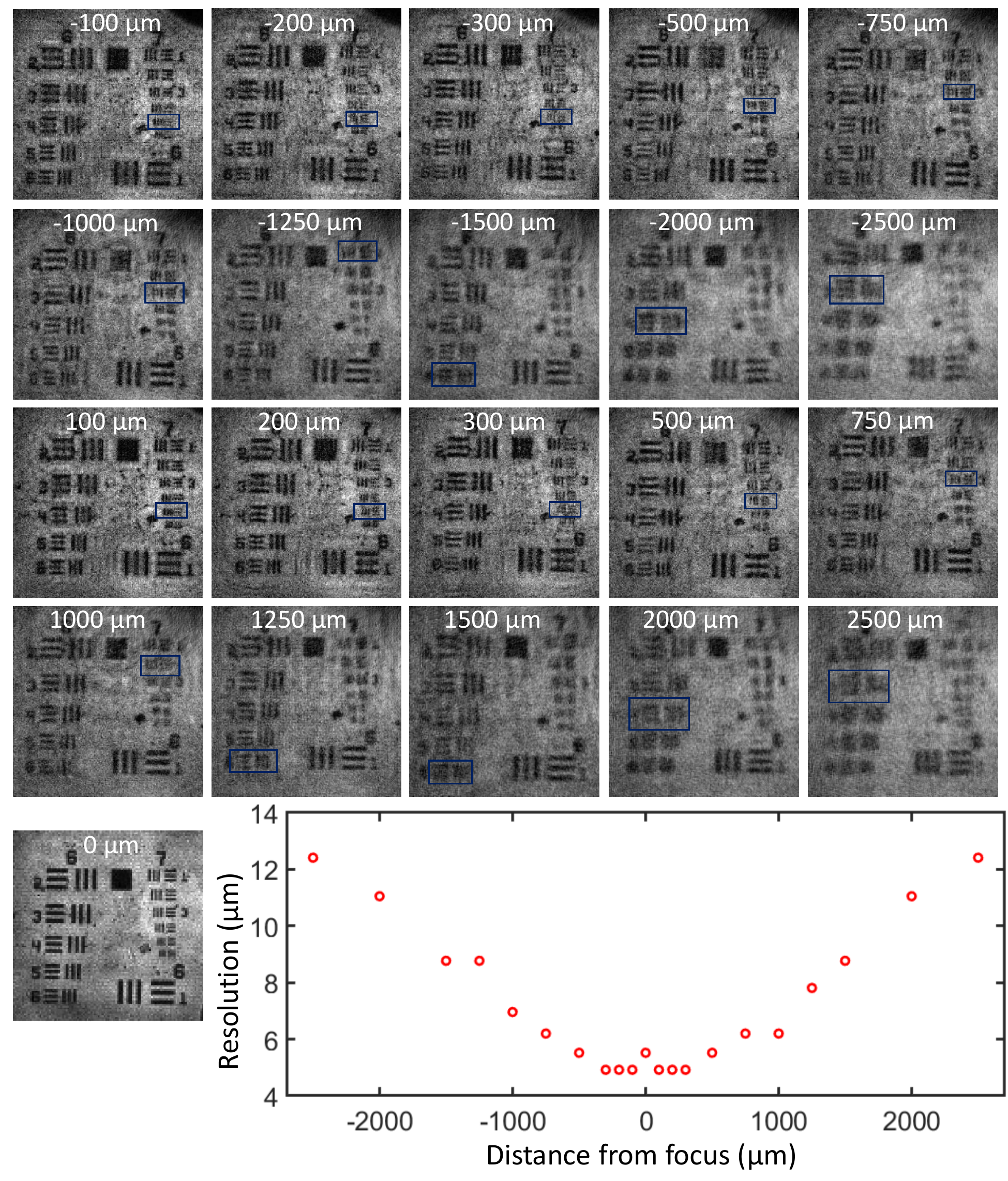}
    \caption{Full dataset of Fig.~3 of the manuscript showing digitally refocused images of groups 6 and 7 of the 1951 USAF resolution target placed at various distances from the objective focus. The smallest resolvable features after refocusing are boxed in blue.} 
    \label{Supp1}
\end{figure}

\begin{figure}
    \centering
    \includegraphics[width=1\linewidth]{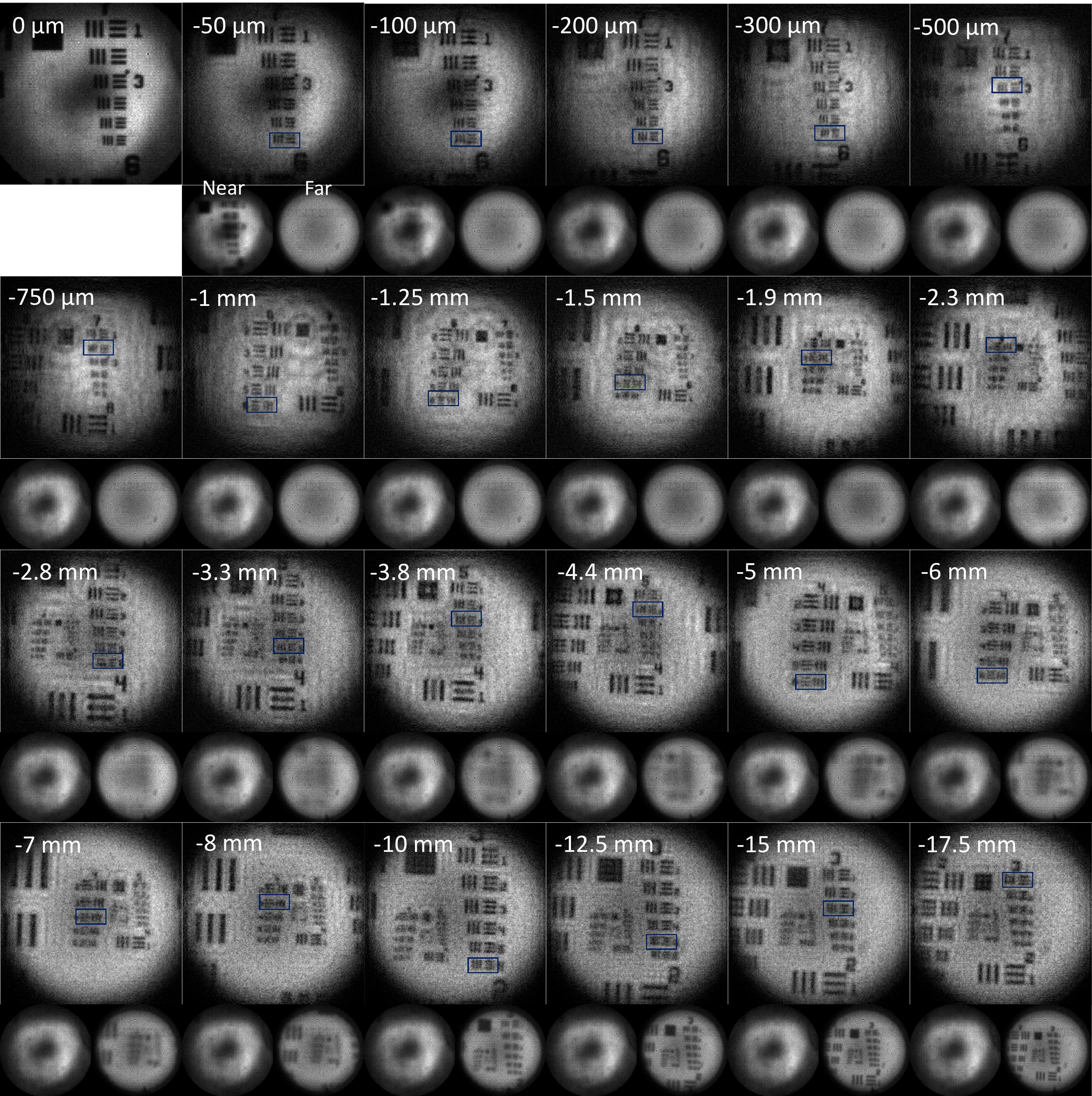}
    \caption{Digitally refocused images of a 1951 USAF resolution target placed at various distances from the objective focus when using a 35\,mm focal length condenser lens. The centred resolution target group from 0 to -750\,$\mu$m away from the focus is group 7, -1\,mm to -2.3\,mm is group 6, -2.8\,mm to -4.4\,mm is group 5, -5\,mm to -7\,mm is group 4 and -10\,mm to -17.5\,mm is group 3. The smallest resolvable features after refocusing are boxed in blue. Images of the near field crystal plane, imaged by the signal beam, and far field crystal plane, imaged by the idler beam, before refocusing are shown below each refocused image. } 
    \label{Supp2}
\end{figure}

\begin{figure}
    \centering
    \includegraphics[width=1\linewidth]{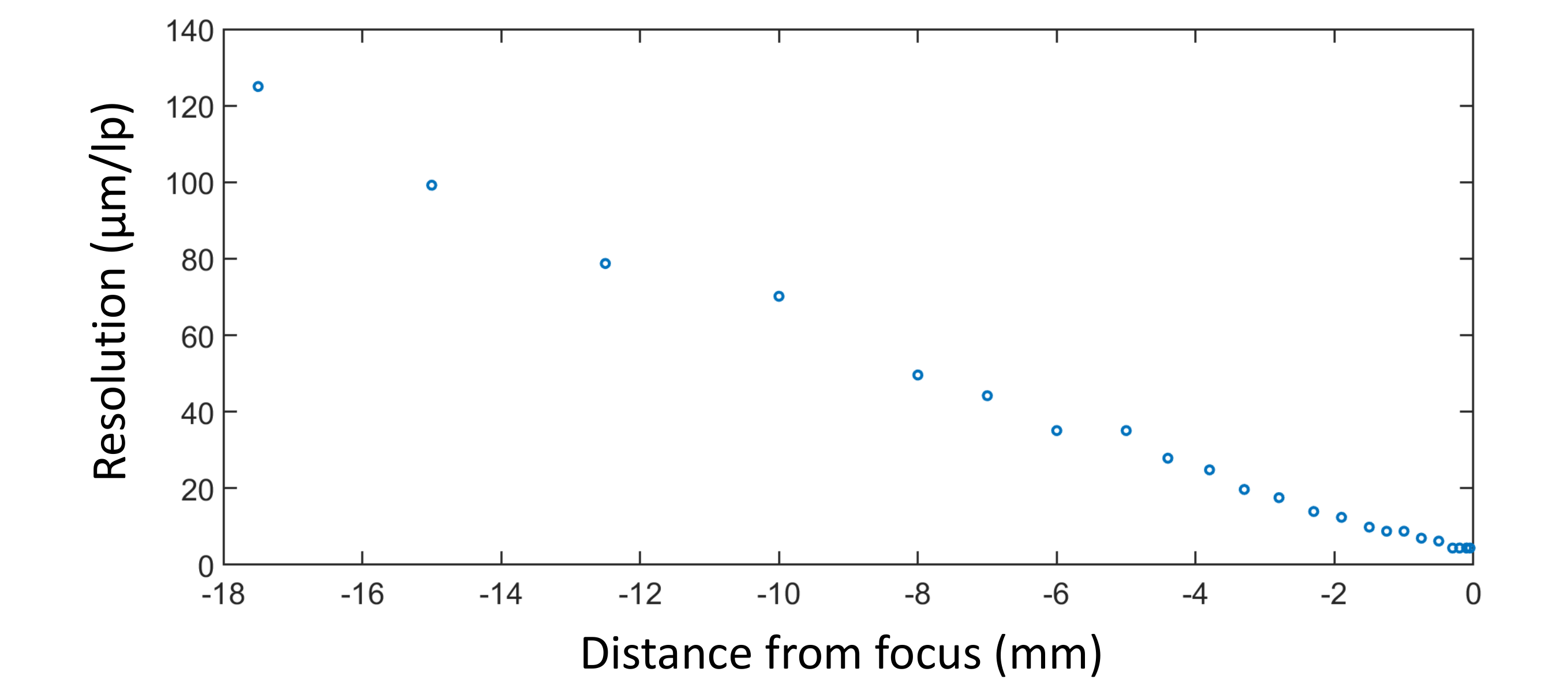}
    \caption{Plot showing the smallest resolvable features vs. the distance from focus for data shown in Fig.~\ref{Supp2}.} 
    \label{Supp3}
\end{figure}

\end{document}